\documentclass[12pt,preprint]{aastex}

\shorttitle{The Emerging Features of BMRs} \shortauthors{Song \&
Feng}

\begin{document}

\title{The Emerging Features of Bipolar Magnetic Regions during Solar Minima}
\author{Wenbin Song and Xueshang Feng}
\affil{State Key Laboratory for Space Weather, Center for Space
Science and Applied Research, Chinese Academy of Sciences, Beijing
100080, China} \email{wbsong@spaceweather.ac.cn}

\begin{abstract}

Solar magnetic synoptic charts obtained by NSO/Kitt Peak and
SOHO/MDI are analyzed for studying the appearance of bipolar
magnetic regions (BMRs) during solar minima. As a result, we find
the emergence of long-lived BMRs has three typical features. (1)
BMRs' emerging rates of the new cycles increase about 3 times
faster than those of the old cycles decrease. (2) Two consecutive
solar cycles have an overlapping period of near 10 Carrington
rotations. During this very short overlapping time interval, BMRs
of two cycles tend to concentrate in the same longitudes. (3)
About 53\% BMRs distribute with a longitudinal distance of 1/8
solar rotation. Such phenomenon suggests a longitudinal mode of
$m=8$ existing during solar minima.

\end{abstract}


\section{Introduction}

New magnetic flux of the sun always appears as a bipolar magnetic
region (BMR) from below in form of two distinct types. One is the
emerging flux region (EFR), the other is the ephemeral region
(ER). EFR is an important ingredient in active region fields which
sometimes can be seen as large-scale patterns, e.g. active
longitudes defined as a sequence of active regions appearing in
the preferred longitudinal bands (Gaizauskas et al. 1983). ER is
relatively small and short-lived (about 1 day), thus it
distributes more randomly. In this paper, using solar magnetic
synoptic charts we investigate patterns in emergence of the
long-lived (at least 1 rotation) BMRs during solar activity
minima. The emergence of BMRs during solar minima is a very
classical topic, addressed by many people, e.g. Howard \& LaBonte
1981, Howard 1989, Wang \& Sheeley 1989, Harvey \& Zwaan 1993, and
many others. Differing from former works,  here  more attention is
paid to the variability of BMRs' emerging rate and the typical
longitudinal distance among them. It is believed that such
information can provide hints for understanding the origin of
solar magnetic fields.

\section{Data Processing}

As the main source of data, we have used NSO/Kitt Peak solar
magnetic synoptic charts and SOHO/MDI synoptic charts. For our
purpose we choose three time intervals of solar minima: Carrington
rotation (CR) 1754-1793, CR1891-1930 and CR2035-2049. During the
former two time intervals we use Kitt Peak synoptic charts and
during the last time interval, we use SOHO/MDI charts since Kitt
Peak charts are not available after CR2007. When dealing with the
high resolution SOHO/MDI charts, we resized the grid from
$3600\times1080$ to the Kitt Peak's grid $360\times180$.

To identify BMRs in each solar magnetic synoptic chart, we set
three criteria. (1) A BMR should possess strong magnetic fields.
Here all pixels with absolute value lower than 20 Gauss are set to
zero (see an example in Figure 1) because they are thought to be
related to the quiet sun and network magnetic fields (Rabin et
al., 1991). (2) BMR's two polarities should distribute closely and
weight balanced;  (3) It should do not exist in the former CR and
leave an obvious track in the following CR. Therefore the BMRs
chosen for every CR should be new and long-lived. Also for this
reason, the statistic of BMRs' distribution won't be affected by
the evolution of the relatively random ERs. After finding a BMR,
we outline its region and calculate its absolute flux gravity
center. In order to cut down the latitudinal measurement errors,
we have used a linear interpolation method to re-map synoptic
charts' sine latitude into equal latitude. As shown in Figure 2,
all BMRs found during CR1906-1925 are signed together by circles.
The center of circle marks the BMR's flux gravity center (or BMR's
position).

\section{Results and Discussion}

Figure 3 depicts the latitudinal distribution of solar emerging
BMRs during the three chosen time intervals (CR1754-1793,
CR1891-1930 and CR2035-2049). According to the Sp\"{o}rer law we
can easily draw the separatrix between the old and new solar
cycles. Here we can see clearly that there is a very special BMR
(see the arrow in Figure 3). Following the discussion of de Toma
et al. (2000), we also think this BMR belongs to cycle 23 and just
emerged much earlier.

\subsection{BMRs' Emerging Rate}

BMRs' emerging rates (the number of new BMRs per CR) during each
solar cycles are computed. The result is shown in the left panel
of Figure 4 where the solid lines indicate BMRs of the old solar
cycles and the dashed lines indicate BMRs of the new solar cycles.
Using a least-square linear fit, we find the BMRs' emerging rates
during the end of cycles 21, 22 descend at an average rate of
$0.052\pm0.016$ $CR^{-2}$ and $0.089\pm0.020$ $CR^{-2}$, while the
emerging rates ascend at a rate of $0.168\pm0.059$ $CR^{-2}$ and
$0.225\pm0.077$ $CR^{-2}$ during the beginning of cycles 22, 23
(regardless of the special BMR mentioned in the above paragraph).
Therefore BMRs' emerging rates of new cycles increase about 3
times faster than those of old cycles decrease. The right panel of
Figure 4 shows the sunspot number during the same time intervals.
We find the sunspot curves in the right panel are very similar to
the BMR curves in the left panel. Some local differences mainly
come from the fact we only compute the new BMRs during each CR.

Figure 3 and Figures 4a, 4b show that two consecutive solar cycles
overlap about 10 CRs. During such overlapping period the BMRs'
emerging rate gets  the least. Table 1 lists the longitudes of
BMRs occurring during the overlapping time interval of cycles 21,
22 and cycles 22,  23. From this we can find that most BMRs of the
old and new cycles tend to concentrate in the same longitudes.
Bumba et al. (2000) also found this phenomenon and suggested this
would be due to  the magnetic flux of the new cycle
 induced by the action of the old cycle magnetic flux.

\subsection{Longitudinal Distance among BMRs}

We mainly study the longitudinal distance ($d=|l_A-l_B|$, $l$
means the BMR's Carrington longitude) between every two BMRs of
the same cycle, emerging in the same CR or $\pm1$ CR
($|\vartriangle{t}|\leq1$ $CR$, $t$ means the BMR's emerging
time), and with a latitudinal distance of no more than $8^{\circ}$
($|\vartriangle{\theta}|\leq8^{\circ}$, $\theta$ means the BMR's
latitude). Only two BMRs satisfying such three conditions can be
called one BMR pair. Here we totally find 155 qualified BMR pairs.
Due to the spherical solar surface, each BMR pairs' longitudinal
distance $d$ should be also regarded as $360^{\circ}-d$ (the unit
of $d$ is $1^{\circ}$). Therefore, we add  the number of BMR pairs
with a distance $d$ and $360^{\circ}-d$. The final result is shown
by the dashed line in Figure 5 (for the bilateral symmetry, we
just draw the former half $d\in[0^{\circ},180^{\circ}]$). With a
measurement of its smoothing effect (see the solid line in Figure
5), we find 82 (or 53\%) BMR pairs to distribute near (peak width
at half-height, $\sigma^{\pm}\in[4^{\circ},11^{\circ}]$, see
Figure 5) four typical longitudinal distances: $d=49^{\circ}$,
$d=95^{\circ}$, $d=136^{\circ}$, and $d=178^{\circ}$. There is
another distinct peak located at around $d=20^{\circ}$. Such $d$
is so small that we think it might originate from BMRs' mutual
action (e.g. while a BMR emerging within another existing BMR,
they may compete for place with each other) or the differential
rotation. In order to further check up our result, we have carried
out similar study for the new/old solar cycle and the
northern/southern hemisphere separately, only to find that the
peak positions are the same.

Very regular four peak positions, $d=49^{\circ}$, $d=95^{\circ}$,
$d=136^{\circ}$, and $d=178^{\circ}$, are   separated up to
multiples of $45^{\circ}$. This phenomenon indicates that most BMRs
tend to distribute with a longitudinal distance of 1/8 solar
rotation. We think such rule should be related to the quantified
distribution of solar magnetic field. Gilman and Dikpati (2000) ever
showed the quantified property of active regions by simulating the
pattern of low-order longitudinal modes. Here a mode of $m=8$ would
be more applicable during solar minima. Song and Wang (2005) found
that about 55\% solar strong magnetic fields can be represented by
two longitudinal modes of $m=5,6$. The different modes during solar
minima suggest that all modes are indeed varying with solar cycle.

\section{Conclusions}

In this paper we have studied the features of the emergence of
long-lived BMRs during solar minima and can draw the following
conclusions. (1) BMRs' emerging rates of the new cycles increase
about 3 times faster than those of the old cycles decrease. (2)
Two consecutive solar cycles have an overlapping period of about
10 CRs. During this very short overlapping time interval, BMRs of
two cycles tend to concentrate in the same longitudes. (3) About
53\% BMRs distribute with a longitudinal distance of 1/8 solar
rotation. This suggests a longitudinal mode of $m=8$ existing
during solar minima.

\acknowledgments

We thank anonymous referee for valuable remarks and suggestions
that resulted in improvement of the paper. This work is jointly
supported by National Natural Science Foundation of China
(40621003, 40536029, and 40604019), the 973 project under grant
2006CB806304, and the CAS International Partnership Program for
Creative Research Teams.

\clearpage

\begin{figure}
\plotone{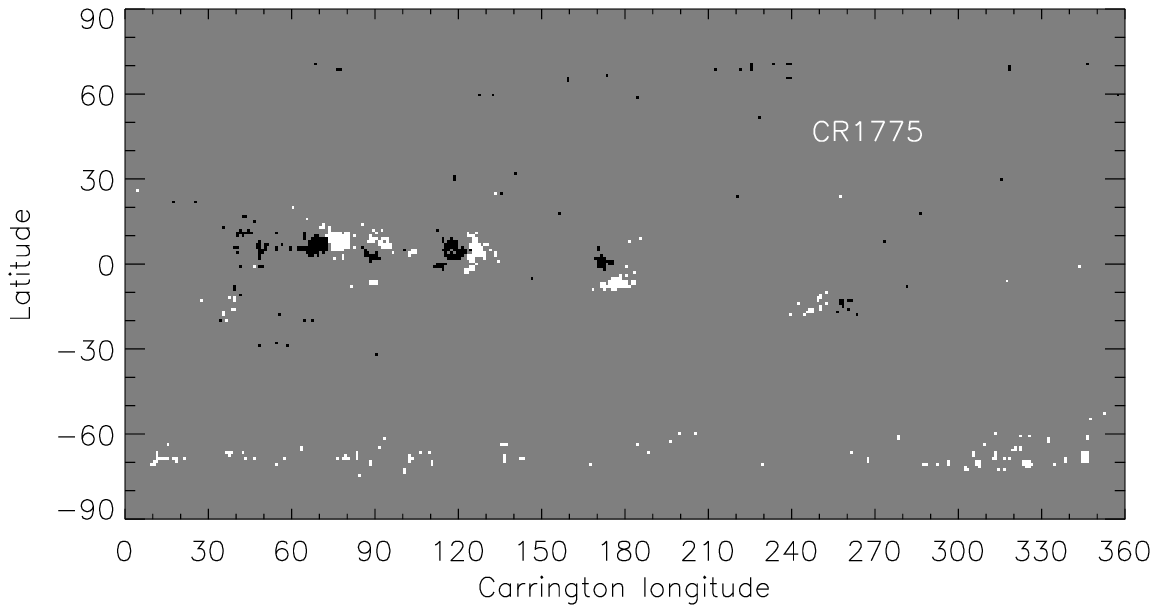} \caption{The NSO/Kitt Peak solar magnetic
synoptic chart of CR1775. Its original sine latitude is re-mapped
into an equal latitude by using a linear interpolation method, and
the pixels with absolute value lower than 20 Gauss are set to
zero. \label{f1}}
\end{figure}

\clearpage

\begin{figure}
\plotone{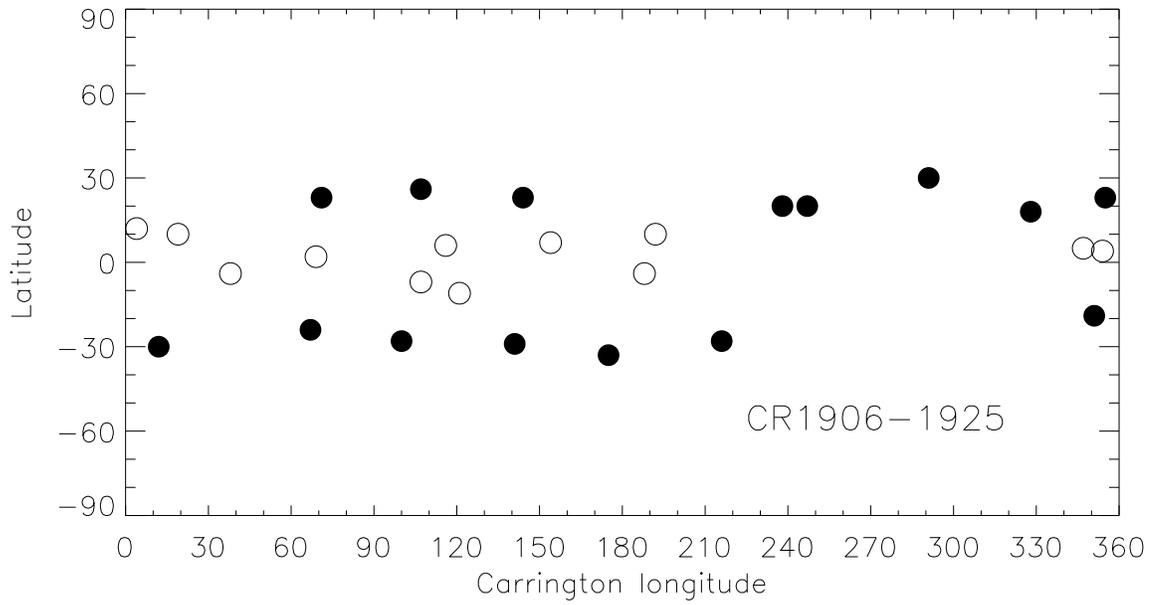} \caption{The surface distribution of solar BMRs
which emerged during CR1906-1925. The center of circles mark BMRs'
flux gravity center. The unfilled and filled circles indicate BMRs
belonging to the old and the new solar cycle, respectively.
\label{f2}}
\end{figure}

\clearpage
\begin{figure}
\plotone{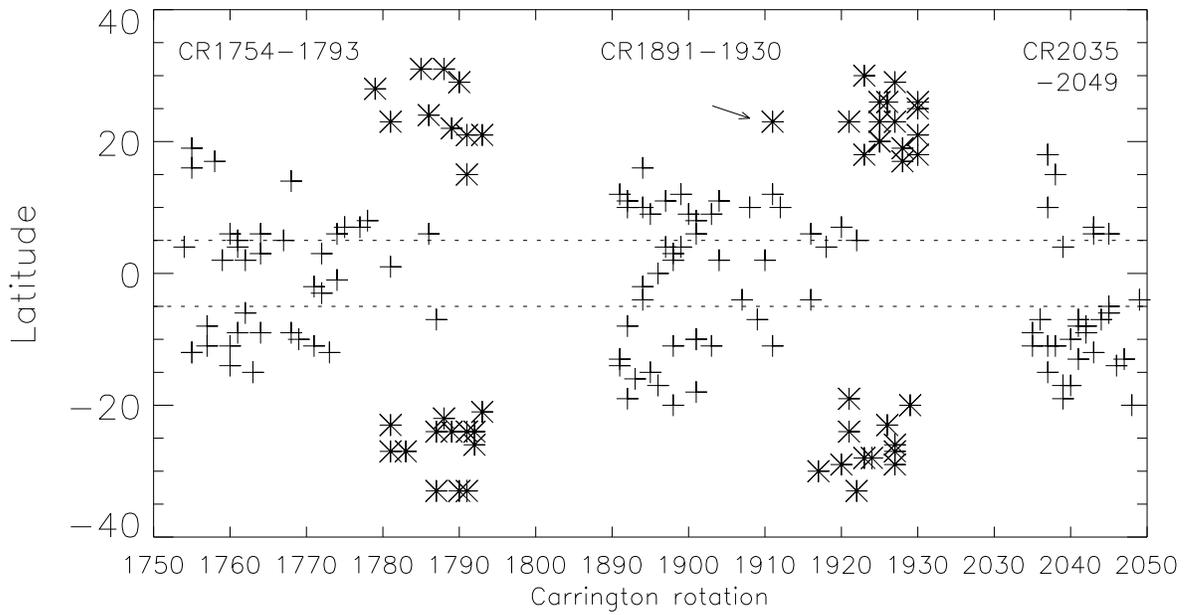} \caption{The latitudinal distribution of all 168
BMRs found during CR1754-1793, CR1891-1930 and CR2035-2049. The
plus signs mark BMRs of old solar cycles and the asterisks mark
BMRs of new solar cycles. The arrow indicates the special BMR of
cycle 23 emerging much earlier. \label{f3}}
\end{figure}

\clearpage
\begin{figure}
\plotone{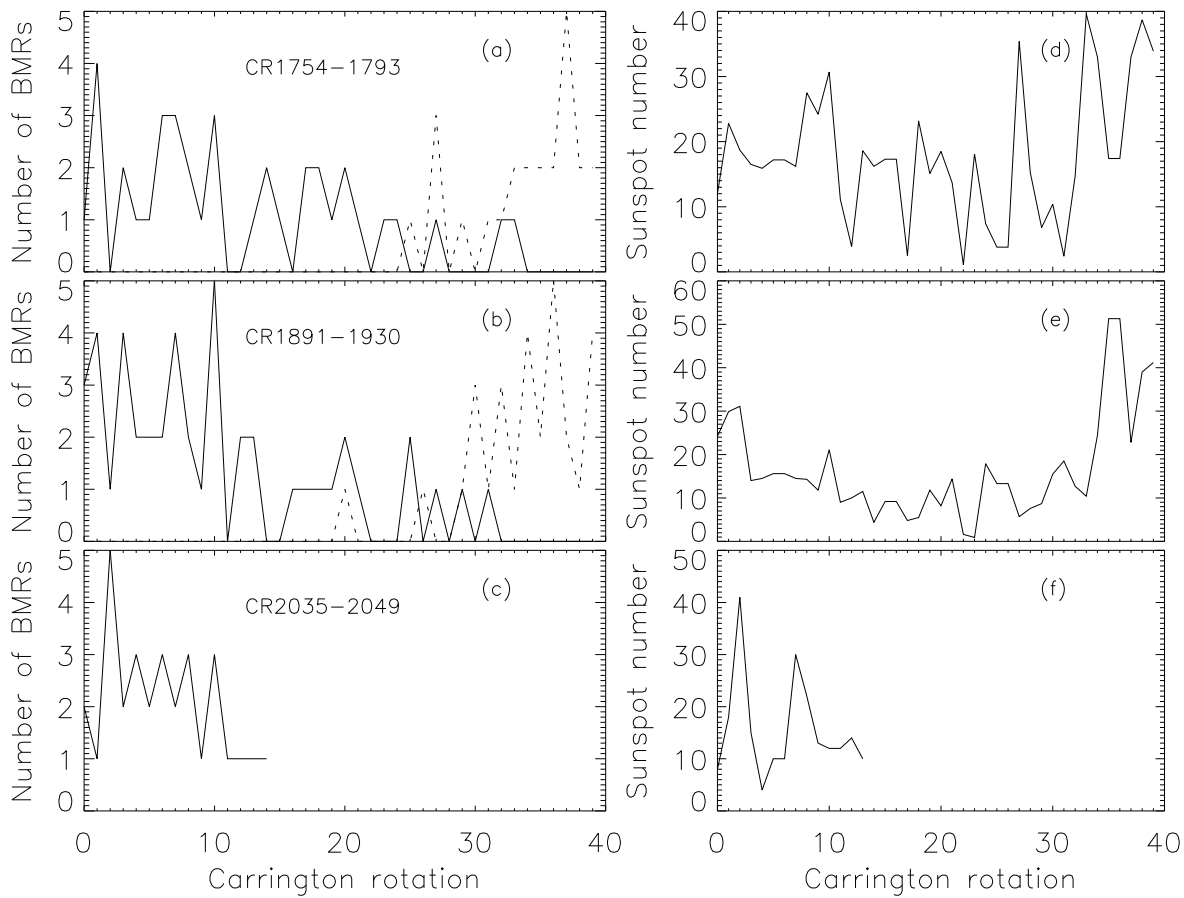} \caption{Left panel: the number of BMRs per CR.
The solid line represents BMRs emerged during the end of an old
solar cycle and the dashed line represents BMRs emerged during the
beginning of a new solar cycle. Right panel: the sunspot number
during the same time intervals, which is published by
NOAA/National Geophysical Data Center. \label{f4}}
\end{figure}

\clearpage
\begin{figure}
\plotone{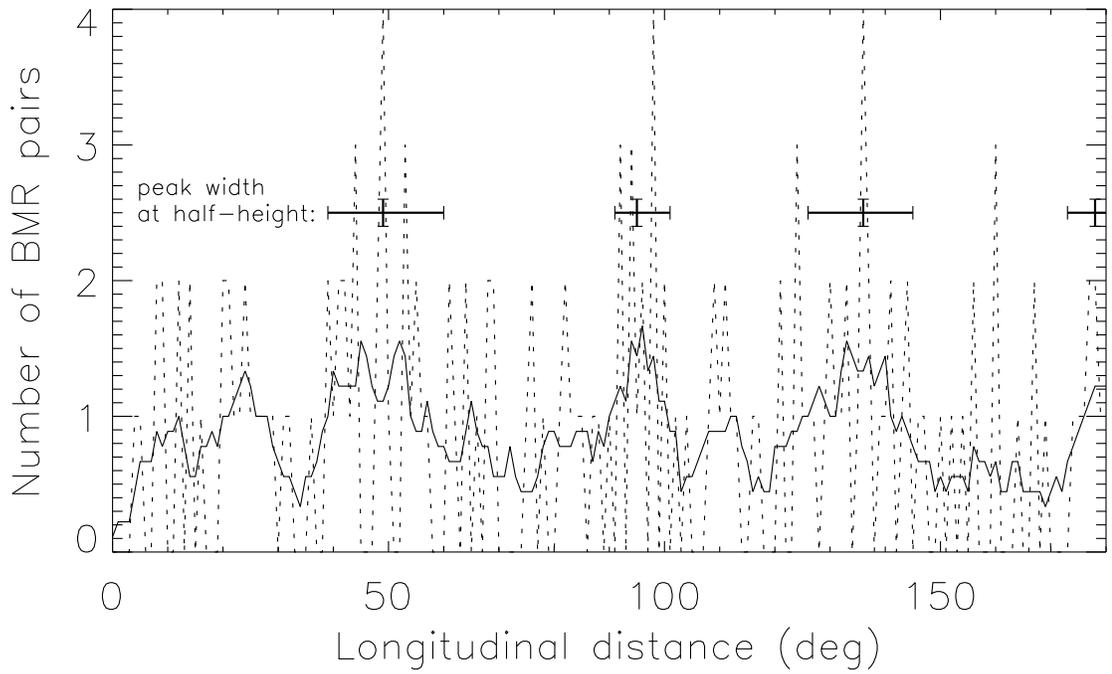} \caption{The dashed line represents the number of
BMR pairs that emerge almost simultaneously
($|\vartriangle{t}|\leq1$ $CR$) and have a similar latitude
($|\vartriangle{\theta}|\leq8^{\circ}$) versus their longitudinal
distances ($d$, bin width: $1^{\circ}$). The solid line is the
dashed line smoothed with a window of $8^{\circ}$. \label{f5}}
\end{figure}

\clearpage
\begin{table}
\begin{center}
\caption{The longitudes ($l$) of BMRs during the overlapping
period of two consecutive solar cycles. All BMRs with a similar
longitude ($\vartriangle{l}\leq25^{\circ}$) are listed in the same
column. From this Table we can see that most BMRs of the old and
new cycles tend to concentrate in the same longitudes.}
 \label{tab:kd}
\begin{tabular}{lccccc}
\hline
    Old cycle - end of cycle 21: & 73 & 130 & & 240 & 270 \\
    New cycle - beginning of cycle 22: & & 134 & 171,180 & 218,221,228 & 269,279,280 \\
\hline
    Old cycle - end of cycle 22: & 69 & 116,121 & 154 & 188 & 347,354,4 \\
    New cycle - beginning of cycle 23: & 67,71 & 100 & 141,144 & 175 & 351,12 \\
\hline
\end{tabular}
\end{center}
\end{table}

\end{document}